\documentclass[
reprint,
amsmath,
amssymb,
aps,
prb,
superscriptaddress
]{revtex4-2}

\usepackage{multirow}
\usepackage{graphicx}
\usepackage{dcolumn}
\usepackage{bm}
\usepackage{hyperref}
\hypersetup{
    colorlinks = true,
    linkcolor = blue,
    citecolor = blue,
    urlcolor = blue,
    pdfborder = {0 0 0},
    pdfusetitle = true
}

\begin{document}

\title{\textbf{Machine learning Hamiltonian enables scalable and accurate defect calculations: \\The case of oxygen vacancies in amorphous SiO$_2$} 
}

\author{Zhenxing Dai}
    \thanks{These authors contribute equally to this work.}
    \affiliation{
    Key Laboratory of Computational Physical Sciences (MOE), Fudan University, Shanghai 200433, China
    }
    \affiliation{Department of Physics, Fudan University, Shanghai 200433, China}
\author{Yang Zhong}
    \thanks{These authors contribute equally to this work.}
    \affiliation{
    Key Laboratory of Computational Physical Sciences (MOE), Fudan University, Shanghai 200433, China
    }
    \affiliation{Department of Physics, Fudan University, Shanghai 200433, China}
\author{Mingjue Ni}
    \affiliation{
    Key Laboratory of Computational Physical Sciences (MOE), Fudan University, Shanghai 200433, China
    }
    \affiliation{Department of Physics, Fudan University, Shanghai 200433, China}
\author{Menglin Huang}
    \email{Contact author: menglinhuang@fudan.edu.cn}
    \affiliation{
    Key Laboratory of Computational Physical Sciences (MOE), Fudan University, Shanghai 200433, China
    }
    \affiliation{College of Integrated Circuits and Micro-Nano Electronics, Fudan University, Shanghai 200433, China}
\author{\\Hongjun Xiang}
    \affiliation{
    Key Laboratory of Computational Physical Sciences (MOE), Fudan University, Shanghai 200433, China
    }
    \affiliation{Department of Physics, Fudan University, Shanghai 200433, China}
\author{Xin-Gao Gong}
    \affiliation{
    Key Laboratory of Computational Physical Sciences (MOE), Fudan University, Shanghai 200433, China
    }
    \affiliation{Department of Physics, Fudan University, Shanghai 200433, China}
\author{Shiyou Chen}
    \email{Contact author: chensy@fudan.edu.cn}
    \affiliation{
    Key Laboratory of Computational Physical Sciences (MOE), Fudan University, Shanghai 200433, China
    }
    \affiliation{College of Integrated Circuits and Micro-Nano Electronics, Fudan University, Shanghai 200433, China}

\begin{abstract}

Point defects critically influence the properties of materials and devices, yet density functional theory (DFT) remains computationally demanding for defect supercell calculations.
Machine learning interatomic potentials (MLIPs) offer high efficiency but require extensive datasets. MLIPs trained only on defect configurations in small supercells exhibit systematic energy errors in larger supercells, demonstrating limited transferability.
Here, we present a machine learning Hamiltonian (MLH) model-based method for calculating total energies and atomic forces in defect supercells with linear-scaling computational cost, enabling efficient structural relaxation and accurate formation energy predictions.
We take oxygen vacancies in amorphous SiO$_2$ as an example and train the MLH model on defect configurations in 95-atom supercells, with the training data derived from 120 self-consistent field calculations and 12 structural relaxations. The MLH model enables efficient structural relaxations for host (defect-free) and defect systems in larger supercells, avoiding the systematic energy errors observed in MLIPs.
The cancellation of energy errors between host and defect systems yields accurate formation energy predictions, with deviations from DFT below 50~meV.
The proposed method holds significant potential for defect simulations in complex materials.

\end{abstract}

\maketitle

\section{INTRODUCTION}
 
Amorphous materials play a vital role in microelectronic and optoelectronic devices, yet their performance is strongly affected by point defects~\cite{stoneham2001theory,baliga2018fundamentals}. Oxygen vacancies (V$_\text{O}$) in amorphous SiO$_2$ (a-SiO$_2$), for instance, introduce defect states that act as carrier traps, lead to bias temperature instability, and enhance leakage current, degrading device performance and reliability~\cite{grasser2011paradigm,grasser2012stochastic,goes2018identification,guo2025si}.
While experimental techniques such as deep-level transient spectroscopy and electron paramagnetic resonance are widely used for defect characterization, revealing the microscopic nature of defects in materials remains challenging owing to their diversity and structural complexity.
Currently, ab initio methods based on density functional theory (DFT) have become essential tools for investigating point defects, with calculations typically employing supercells with hundreds of atoms.
However, the self-consistent field (SCF) iteration in DFT is computationally expensive for defect supercell calculations, severely limiting the scale and efficiency of simulations.
To accurately capture the statistical behavior of defects in amorphous materials, extensive sampling of defect sites is required, which further raises the computational cost.
Therefore, developing a method that maintains first-principles accuracy while efficiently predicting defect properties holds significant importance.

To achieve this goal, combining machine learning methods with DFT has emerged as a promising approach.
In recent years, machine learning interatomic potentials (MLIPs)~\cite{batatia2022mace,deng2023chgnet,batzner20223,musaelian2023learning}
have been increasingly applied in the study of point defects, including high-throughput defect screening~\cite{berger2025screening}, 
potential energy surface exploration~\cite{jiang2024machine,mosquera2024machine},
investigations of defect formation and migration mechanisms~\cite{shimizu2022using,wu2023oxygen,mosquera2025point,chen2025simulating},
and simulations of defect evolution during the irradiation damage process
\cite{song2023neural}.

However, the application of MLIPs to point defects still faces challenges.
First, training MLIPs requires large datasets, which rely on extensive first-principles calculations, rendering dataset construction highly challenging.
For example, Mosquera-Lois \textit{et al.}~\cite{mosquera2024machine} built a dataset containing about 4,000 DFT data points for Te$_\text{i}^{+1}$ in CdTe to investigate the contribution of vibrational entropy to the defect formation free energy.
Shimizu \textit{et al.}~\cite{shimizu2022using} collected over 20,000 DFT data points for the GaN host and nitrogen vacancies in different charge states to predict defect formation energies, phonon spectra, and the temperature dependence of Grüneisen parameters and transition levels.
Wu \textit{et al.}~\cite{wu2023oxygen} sampled 8,544 configurations of V$_\text{O}$ to study their diffusion mechanism on the rutile TiO$_2$(110) surface.
These examples show that dataset construction can be time-consuming even for one defect species.

Another practical challenge is the transferability of MLIPs for defect properties in varying supercells. Mosquera-Lois \textit{et al.}~\cite{mosquera2024machine} found that an MLIP trained on Te$_\text{i}^{0}$ defects in small CdTe supercells exhibited systematic energy errors for Te$_\text{i}^{0}$ defects in larger supercells, which stemmed from the MLIP's inaccurate estimation of host-atom contributions. They suggested that, to address this issue, the MLIP should be trained on both host and defect supercells. 

Moreover, predicting defect formation energies with MLIPs is complicated by the fact that training solely on defect-containing supercells leaves the host description inaccurate, forcing one to still rely on DFT for host properties. A related issue was highlighted by Wang \textit{et al.}~\cite{wang2024perovsdopants}. They found that the foundation model MACE-MP pretrained on the MPtrj dataset~\cite{deng2023chgnet, batatia2025foundation} performed well on the MPtrj subset but struggled with doped perovskites. Fine-tuning it with a new dopant dataset improved predictions for doped systems but resulted in catastrophic forgetting on the MPtrj dataset. Consequently, addressing these transferability and forgetting issues requires vastly expanding the training sets, which further raises computational costs. Developing methods that achieve high accuracy and transferability with modest datasets remains a critical goal.

Given the challenges faced by MLIPs, we redirect our focus toward machine learning Hamiltonian (MLH) models~\cite{li2022deep,zhong2023transferable,zhong2024universal,haldar2025gears}. 
Our recent research demonstrated that MLH models could precisely predict the band structures for both host and defect systems with a relatively small dataset, exhibiting excellent generalization performance.
For example, an MLH model trained on a dataset of 30 silicon polymorph configurations can not only accurately predict the band structures for silicon polymorphs outside the training set, but also those for dislocation defects in the silicon supercell~\cite{zhong2023transferable}. More recently, Haldar \textit{et al.}~\cite{haldar2025gears} also successfully applied the MLH model to predict electronic band structures for defective WSe$_2$ interfaced with Ni-doped amorphous HfO$_2$. These studies demonstrate the validity of the MLH methods for electronic structure simulations of defects. However, a common limitation of these studies is that they cannot obtain the total energy and atomic forces from the MLH, and therefore still require atomic structural relaxation using expensive DFT calculations to obtain the correct atomic structures of defects. This limitation hinders the practical application of MLH methods to new materials with unknown atomic structures. To the best of our knowledge, the direct derivation of total energy and atomic forces from an MLH has not yet been achieved; consequently, the application of MLH methods for defect structural relaxation and formation energy prediction remains unexplored.

In this study, we developed an MLH-based method for computing the total energies and atomic forces in defect supercells with linear scaling computational cost, enabling efficient and accurate structural relaxation and formation energy prediction.
Here, we take V$_\text{O}$ defects in a-SiO$_2$ as an example 
and train the MLH model only on defect configurations in 95-atom supercells, with the training data obtained from 120 SCF calculations and 12 structural relaxations. The trained MLH model enables accurate structural relaxations of a-SiO$_2$ hosts and V$_\text{O}$ defects in larger supercells, yielding relaxed configurations and the corresponding total energies in good agreement with DFT results, free from the systematic energy errors observed in MLIPs.
During formation energy calculations, the total energy errors of host and defect systems partially cancel each other, reducing the defect formation energy error to below 50 meV relative to DFT.
Therefore, the proposed approach can significantly enhance the efficiency of large-scale defect simulations in complex materials while maintaining DFT accuracy.

\section{METHODS}

\begin{figure*}[ht] 
    \centering
    \includegraphics[scale = 0.69]{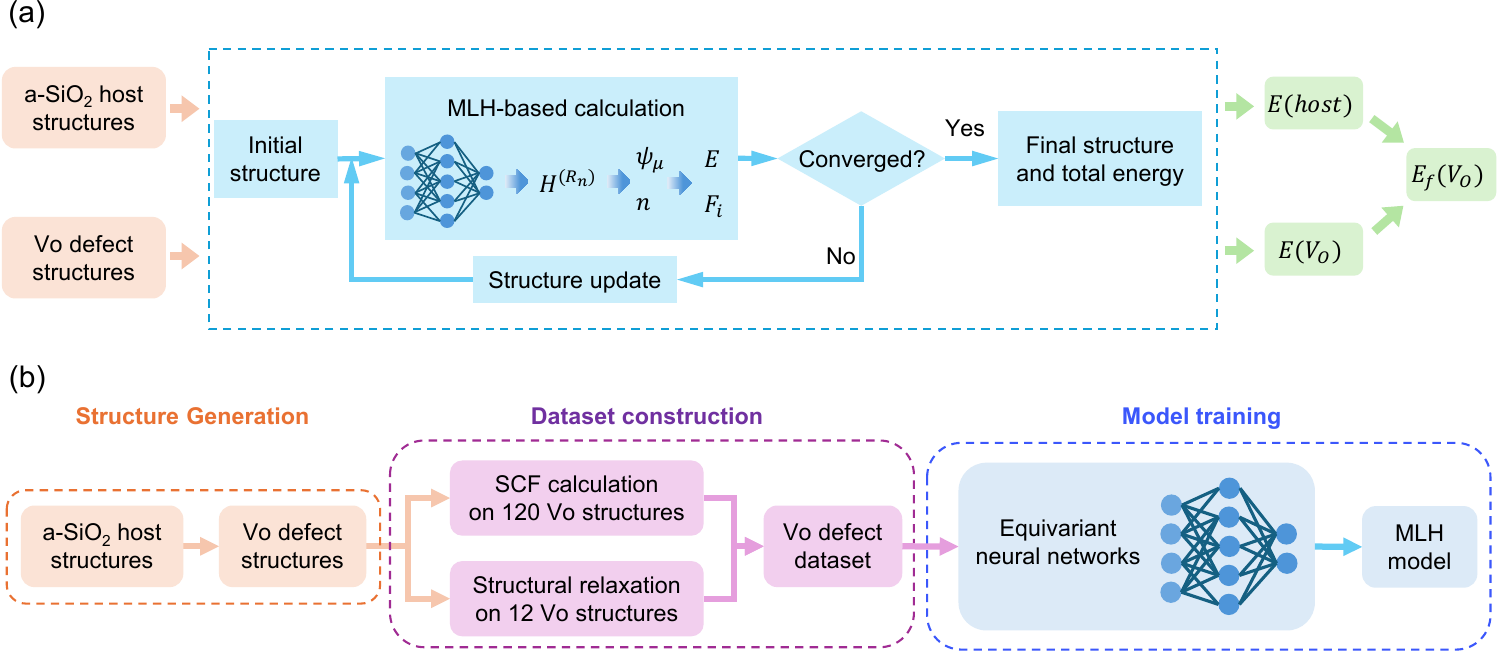}
    \caption{(a) Workflow of the MLH-based method for structural relaxation and defect formation energy calculations. (b) Schematic diagram of the machine learning processes of the MLH model. }
    \label{MLH_process}
\end{figure*}

\subsection{Defect formation energy and DFT details}

In this work, we focus on the formation energy of neutral defects, which 
can be typically calculated as follows~\cite{freysoldt2014first,huang2022dasp}:
\begin{equation}
    E_f(\text{defect}) = E(\text{defect}) - E(\text{host}) + \sum_i n_i (E_i+\mu_i)
\end{equation}
Here, $E(\text{defect})$ denotes the total energy of the supercell with a relaxed defect, and $E(\text{host})$ is the total energy of the host for the same supercell without the defect.
$\mu_i$ denotes the chemical potential of the element $i$, referenced to the single-atom energy $E_i$ of the corresponding elemental solid/gas at the stable phase, and $n_i$ is the number of atoms removed from ($n_i > 0$) or added to ($n_i < 0$) the supercell to form the defect.
For V$_\text{O}$ defects in a-SiO$_2$, we consider the defect formation energy under the oxygen-rich condition, hence $\mu_\text{O} = 0$. 
As illustrated in Fig.~\ref{MLH_process}(a), to predict defect formation energies using the MLH model, we perform structural relaxations for a-SiO$_2$ hosts and V$_\text{O}$ defects. At each relaxation step, the Hamiltonian is predicted with the MLH model, followed by computation of the total energies and atomic forces. The relaxation proceeds until the atomic forces drop below 20 meV/Å. Subsequently, the formation energy is calculated using the defect total energy $E(V_\text{O})$ and the host total energy $E(\text{host})$ predicted by the MLH model, with $E(\text{O}_2)$ still obtained via conventional DFT calculations.

All DFT calculations are performed using the OpenMX ab initio package~\cite{ozaki2003variationally,ozaki2005efficient}, which is based on norm-conserving pseudo-potentials and pseudo-atomic localized basis functions. 
We use the generalized gradient approximation (GGA) with the Perdew-Burke-Ernzerhof (PBE) functional to treat the exchange–correlation energy~\cite{perdew1996generalized}.
Si-2s2p1d and O-2s2p1d pseudo-atomic orbitals are used as the basis to expand wavefunctions.
The energy cutoff is set at 200 Ry, with a convergence criterion of $1.0 \times 10^{-7}$ Hartree. Details of the k-point meshes used for supercells of different sizes are provided in the Supplemental Material.

\subsection{Total energy and atomic forces}

In this section, we describe how the total energy and atomic forces of the system are computed using the Hamiltonian predicted by the MLH model.
In DFT, the ground-state properties of a system are described by Kohn-Sham (KS) equations.
In the atomic orbital basis set, the KS equation at a given $k$-point in reciprocal space is expressed as a generalized eigenvalue problem~\cite{ozaki2003variationally}:
\begin{equation}
    H^{(\mathbf{k})}c^{(\mathbf{k})}_{\mu}=\varepsilon^{(\mathbf{k})}_{\mu}S^{(\mathbf{k})}c^{(\mathbf{k})}_{\mu}
\end{equation}
where
$c^{(\mathbf{k})}_{\mu}$ is the vector of coefficients to expand the $\mu$-th wavefunction, $\varepsilon^{(\mathbf{k})}_{\mu}$ is the corresponding eigenvalue, 
$H^{(\mathbf{k})}$ and $S^{(\mathbf{k})}$
are the Hamiltonian and overlap matrix in the reciprocal space, obtained by Fourier transform of the real-space Hamiltonian $H^{(\mathbf{R_n})}$ and overlap matrix $S^{(\mathbf{R_n})}$:
\begin{equation}
\begin{aligned}
    H^{(\mathbf{k})}_{i\alpha,j\beta}
    & = \sum_{\mathbf{R_n}}e^{i\mathbf{R_n}\cdot \mathbf{k}} H^{(\mathbf{R_n})}_{i\alpha,j\beta} \\
    & = \sum_{\mathbf{R_n}}e^{i\mathbf{R_n}\cdot \mathbf{k}}
    \langle \phi_{i\alpha}(\mathbf{r}-\tau_i) | \hat{H}|\phi_{j\beta}(\mathbf{r}-\tau_j-\mathbf{R_n})\rangle
\end{aligned}
\end{equation}
\begin{equation}
\begin{aligned}
    S^{(\mathbf{k})}_{i\alpha,j\beta} 
    &= \sum_{\mathbf{R_n}}e^{i\mathbf{R_n}\cdot \mathbf{k}} S^{(\mathbf{R_n})}_{i\alpha,j\beta} \\
    &= \sum_{\mathbf{R_n}}e^{i\mathbf{R_n}\cdot \mathbf{k}}
    \langle \phi_{i\alpha}(\mathbf{r}-\tau_i) | \phi_{j\beta}(\mathbf{r}-\tau_j-\mathbf{R_n})\rangle    
\end{aligned}
\end{equation}
where $\{\phi_{i\alpha}\}$ are pseudo-atomic orbital basis functions,
$i$ and $j$ label different atoms, $\alpha$ and $\beta$ denote different atomic orbitals, $\tau_i$ and $\tau_j$ are the positions of the corresponding atoms, and $\mathbf{R_n}$ is the shift vector of the periodic image cell.
Here, $H^{(\mathbf{R_n})}$ can be predicted by the MLH model, and $S^{(\mathbf{R_n})}$ can be computed analytically.

Once the expansion coefficients are obtained,
the wavefunction $\psi^{(k)}_{\mu}$ can be constructed as:
\begin{equation}
    \psi^{(\mathbf{k})}_{\mu} = \frac{1}{\sqrt{N}}\sum_{\mathbf{R_n}}e^{i\mathbf{R_n}\cdot \mathbf{k}}
    \sum_{i\alpha}c^{(\mathbf{k})}_{\mu,i\alpha}\phi_{i\alpha}(\mathbf{r}-\tau_i-\mathbf{R_n})
\end{equation}
and the charge density $n(\mathbf{r})$ is then computed as:
\begin{equation}
    \begin{aligned}
        n(\mathbf{r}) & = \frac{1}{V_B}\int_B d^3k \sum_\mu^{occ} \sum_{i\alpha,j\beta}
        \sum_{\mathbf{R_n}}e^{i\mathbf{R_n}\cdot \mathbf{k}} \\
        & ~~~~~~ \times  c^{(\mathbf{k})*}_{\mu,i\alpha}c^{(\mathbf{k})}_{\mu,j\beta}
        \phi_{i\alpha}(\mathbf{r}-\tau_i)  \phi_{j\beta}(\mathbf{r}-\tau_j-\mathbf{R_n})
    \end{aligned}
\end{equation}

The total energy $E$ is given by the sum of the kinetic energy $E_{kin}$, the electron-core Coulomb energy $E_{ec}$, the electron-electron Coulomb energy $E_{ee}$, the exchange-correlation energy $E_{xc}$ and the core-core Coulomb energy $E_{cc}$~\cite{ozaki2005efficient}, where
\begin{equation}
    \begin{aligned}
    E_{kin} &= \frac{1}{V_B}\int_B d^3 k \sum_\mu^{occ}
    \langle \psi^{(\mathbf{k})}_\mu|\hat{T}|\psi^{(\mathbf{k})}_\mu\rangle
    \\
    &= \frac{1}{V_B}\int_B d^3k \sum_\mu^{occ} 
    \sum_{i\alpha,j\beta} \sum_{\mathbf{R_n}}e^{i\mathbf{R_n}\cdot \mathbf{k}}
    c^{(\mathbf{k})*}_{\mu,i\alpha}c^{(\mathbf{k})}_{\mu,j\beta} \\
    & \times\langle \phi_{i\alpha}(\mathbf{r}-\tau_i) | \hat{T}|\phi_{j\beta}(\mathbf{r}-\tau_j-\mathbf{R_n})\rangle ,
    \end{aligned}  
\end{equation}
\begin{equation}
    \begin{aligned}
    E_{ec} &= E_{ec}^{(L)}+E_{ec}^{(NL)} = \int d^3r ~n(\mathbf{r})\sum_i V_i^{(L)}(\mathbf{r}-\tau_i) \\
    & +\int d^3r ~n(\mathbf{r})\sum_i V_i^{(NL)}(\mathbf{r}-\tau_i)
    \end{aligned}  
\end{equation}
Here, $V_i^{(L)}$ and $V_i^{(NL)}$ are the local part and non-local part in the pseudo-potential of atom $i$.
$E_{ee}$ and $E_{xc}$ are functionals of the charge density $n(\mathbf{r})$, and $E_{cc}$ is a repulsive interaction between nuclei and depends only on their positions. Thus, the total energy $E$ can be obtained, and the atomic forces on atom $i$ are given by the partial derivative of the total energy with respect to its atomic coordinates.

\begin{figure*}[htbp]
    \centering 
    \includegraphics[scale = 0.45]{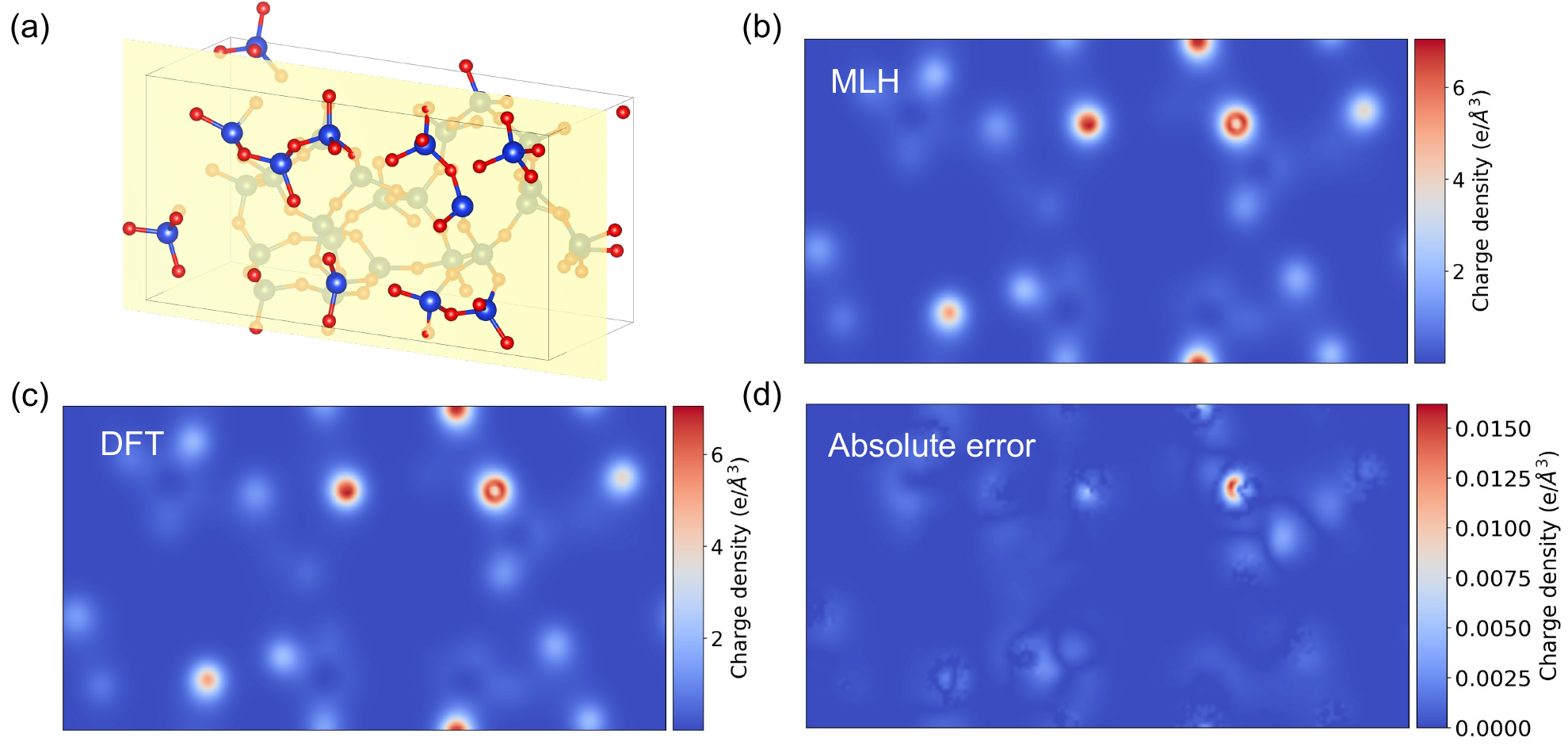}
    \caption{
    (a) The atomic structure of the V$_\text{O}$ defect and the selected plane for charge density comparison. (b) The DFT-calculated and (c) MLH-predicted charge density on the selected plane. (d) The absolute error of MLH-predicted charge density.} 
    \label{MLH_charge}
\end{figure*}

\subsection{Dataset and model training}

Fig.~\ref{MLH_process}(b) illustrates the workflow for training the MLH model, comprising three steps: structure generation, dataset construction, and model training.
In the first step, we employ the Bond Switch Monte Carlo (BSMC) method~\cite{zheng2017effects,zheng2025generalized} to generate a-SiO$_2$ host structures.
The BSMC simulation starts from a 96-atom crystalline SiO$_2$ supercell, maintaining a fixed lattice constant throughout the process.
A disordered structure is obtained after about 5000 melting steps, and the system finally reaches an equilibrium state after a quenching process of up to 13000 Monte Carlo steps, during which the temperature is gradually decreased from 2000 K to 300 K.
Two 96-atom crystalline SiO$_2$ supercells with distinct lattice constants are employed to generate a total of six 96-atom a-SiO$_2$ structures for dataset construction. The lattice constants and radial distribution functions of these amorphous structures are provided in the Supplemental Material. 

In the second step, to construct the dataset, we subsequently select 20 oxygen atom sites in each host to construct V$_\text{O}$ defects and perform SCF calculations on each defect.
Then, two defects per host are selected for DFT structural relaxation until the forces converge below 20 meV/Å. We sample 40 defect configurations along each relaxation trajectory and add them to the dataset.
Thus, the final dataset contains 120 unrelaxed V$_\text{O}$ structures and 480 derived from the relaxation trajectories.

We use HamGNN, a transferable E(3) graph neural network, to train the MLH model~\cite{zhong2023transferable,zhong2024universal}.
In HamGNN, the real-space Hamiltonian matrices are decomposed into on-site and off-site parts, which are obtained by the representation transformation of node features and edge features in the atomic structure. The node features are updated through an equivariant message-passing function in the orbital convolution layer, and the edge features are formed by integrating the atom pair features in the pair interaction layer. The detailed description of the neural network architecture is discussed in Ref.~\cite{zhong2023transferable}.

The MLH model has 3 orbital convolution layers.
The equivariant node features we used are 64$\times$0e + 64$\times$0o + 32$\times$1o + 16$\times$1e + 12$\times$2o + 25$\times$2e + 18$\times$3o + 9$\times$3e + 4$\times$4o + 9$\times$4e. ``64$\times$0e'' means there are 64 channels in each feature part, where features in each channel are O(3) irreducible representations with rotation order $l=0$, and the indices ``e'' and ``o'' stand for even and odd parity, respectively. 
The interatomic distance $\tau_{ij}$ between atom $i$ and its neighboring atom $j$ within the cutoff radius $r_c$ is expanded using a set of Bessel functions:
\begin{equation}
    B(|\tau_{ij}|)=\sqrt{\frac{2}{r_c}}\frac{\sin (n\pi |\tau_{ij}|/r_c)}{|\tau_{ij}|}
\end{equation}
where the index $n = 1,2,\dots,N_b$. We use $N_b = 64$, and $r_c = 26$ Bohr in this work. The relative directions are embedded into spherical harmonic basis functions with a degree of $l_{max}$ = 4, and the irreducible representation is set to 0e + 1o + 2e + 3o + 4e.
The dataset is randomly divided into training, validation and test sets in a ratio of 8:1:1. 
The MAE of Hamiltonian matrix elements is used as the loss function, and the AdamW optimizer is used to optimize the network parameters. The learning rate is reduced from the initial setting of 0.01 to $10^{-6}$. After training, the MAE of the Hamiltonian matrix elements for structures in the test set is 0.26 meV. 

For comparison with the MLH model, an MLIP is trained using Allegro~\cite{batzner20223,musaelian2023learning}. There are 3 tensor product layers in the MLIP with maximum angular quantum number of $l_{max} = 2$, and the cutoff radius is 6\,\AA. We randomly divide the dataset into training and validation sets in a ratio of 9:1. The initial learning rate is set to 0.002, and subsequently decays according to an on-plateau scheduler with a tolerance value of 20 and a decay factor of 0.8. The joint mean squared error of per-atom energies and atomic forces is used as the loss function, and the network parameters are optimized using the Adam optimizer until the learning rate drops to $10^{-5}$. After training, the MAEs of per-atom energies for the defect structures in the training and validation sets are 0.50 meV and 0.58 meV, respectively, and the corresponding force MAEs are 8.3 meV/Å and 20.4 meV/Å, respectively. More details are provided in the Supplemental Material.

All models are trained on a single NVIDIA GeForce RTX 4090 GPU. All subsequent benchmark testing for computational efficiency between the DFT and MLH models is conducted on a single computing node equipped with a 64-core Intel Xeon Platinum 8375C CPU.

\section{RESULTS AND DISCUSSION}

\subsection{The performance of the MLH model}

\begin{figure}[h]
    \centering 
    \includegraphics[scale = 0.44]{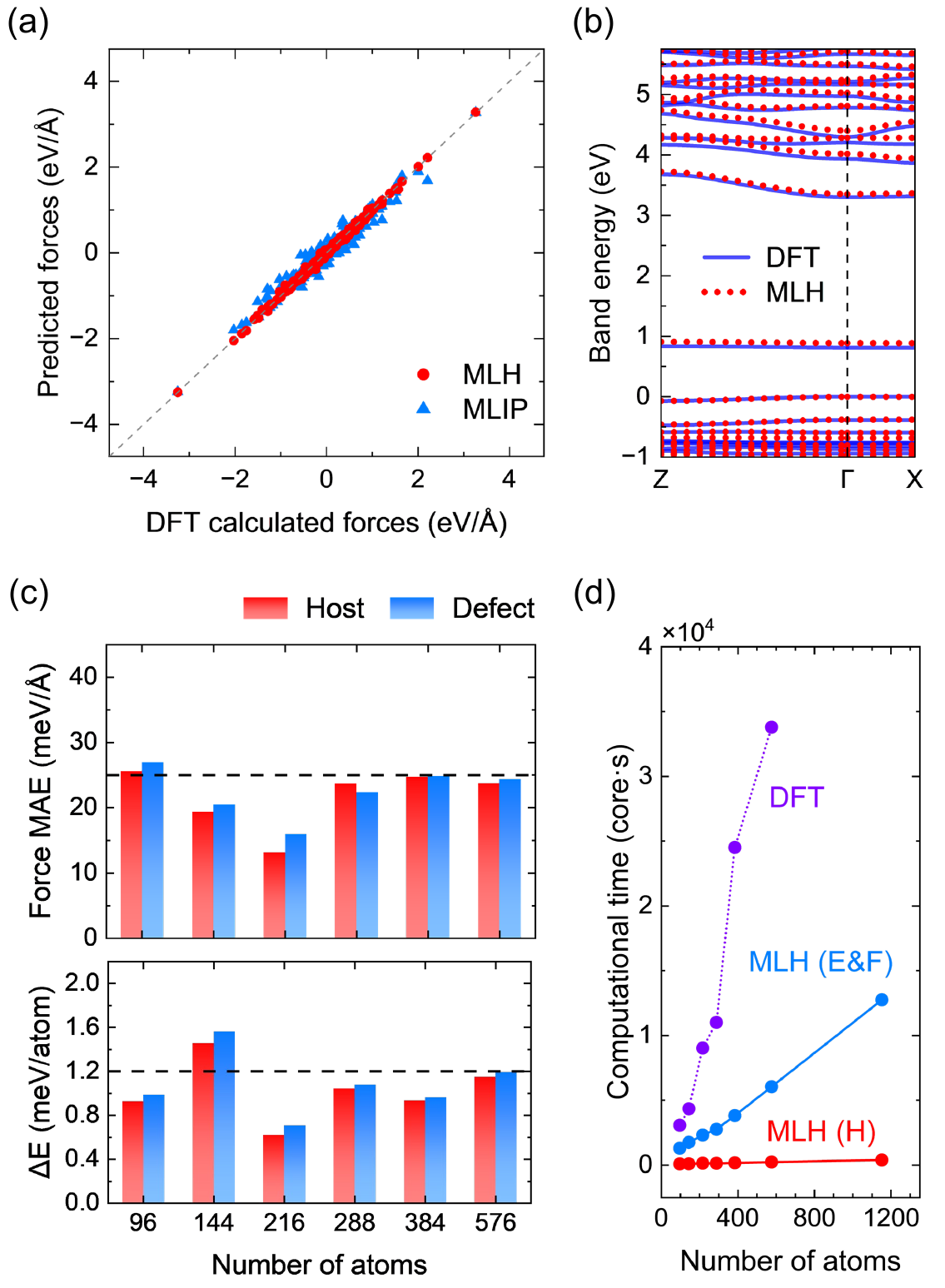}   
    \caption{
    Testing results for a V$_\text{O}$ defect: (a) MLH and MLIP atomic forces vs. DFT; (b) MLH band structure vs. DFT.
    (c) Total energy and atomic force errors of MLH for a-SiO$_2$ hosts and V$_\text{O}$ defects in supercells of varying sizes. The dashed lines correspond to errors of 25 meV/Å for atomic forces and 1.2 meV/atom for total energy.
    (d) Computational time comparison between MLH and DFT. MLH (H): Hamiltonian matrix prediction; MLH (E\&F): total energy and atomic force calculation.}
    \label{MLH_size}
\end{figure}

To assess the accuracy and transferability of the MLH model, we apply it to a V$_\text{O}$ defect in a previously unseen 95-atom supercell. Fig.~\ref{MLH_charge}(a) shows the structure of this defect. We predict the Hamiltonian matrix by the MLH model and obtain a MAE of only 0.72 meV relative to DFT. A scatter plot comparing the MLH‑predicted and DFT-calculated Hamiltonian matrix elements is provided in the Supplemental Material to illustrate the accuracy.
Subsequently, we diagonalize the Hamiltonian matrix to obtain the wavefunctions and derive the charge density, which is not achievable with MLIPs. Compared to DFT, the MAE of the charge density is $3\times10^{-4}$ e/\AA$^3$. We then extract the charge density on the selected plane shown in Fig.~\ref{MLH_charge}(a). The results show that the MLH‑predicted charge density closely overlaps the DFT result with absolute errors below 0.016 e/\AA$^3$, as presented in Fig.~\ref{MLH_charge}(b-d). 

From the predicted charge density, we finally obtain the total energy and atomic forces. Comparison with DFT shows that the MLH-predicted total energy is higher by 94 meV, corresponding to a per-atom energy error of 1 meV/atom, and the MAE of atomic forces is 27 meV/Å, approaching the typical convergence criteria of DFT calculations. In comparison, the errors of the MLIP in total energy and atomic forces are 990 meV and 115 meV/Å, respectively, significantly higher than those of the MLH model. To visualize this contrast, Fig.~\ref{MLH_size}(a) compares the atomic forces predicted by the MLH model and the MLIP against the DFT results. 

Furthermore, we can obtain the band structure from the MLH-predicted Hamiltonian. As shown in Fig.~\ref{MLH_size}(b), the band energies are in close agreement with DFT, with deviations on the order of 0.01 eV. This is also beyond the reach of MLIPs. As depicted in the band structure, the V$_\text{O}$ introduces a deep defect level located approximately 2.5 eV below the conduction band minimum. It should be noted that the apparently narrow bandgap is attributed not only to the intrinsic bandgap underestimation of the PBE functional, but also to the fact that the 0 eV reference corresponds to structurally induced localized tail states rather than the extended valence band edge.

\begin{figure*}[t!]
    \centering 
    \includegraphics[scale = 0.6]{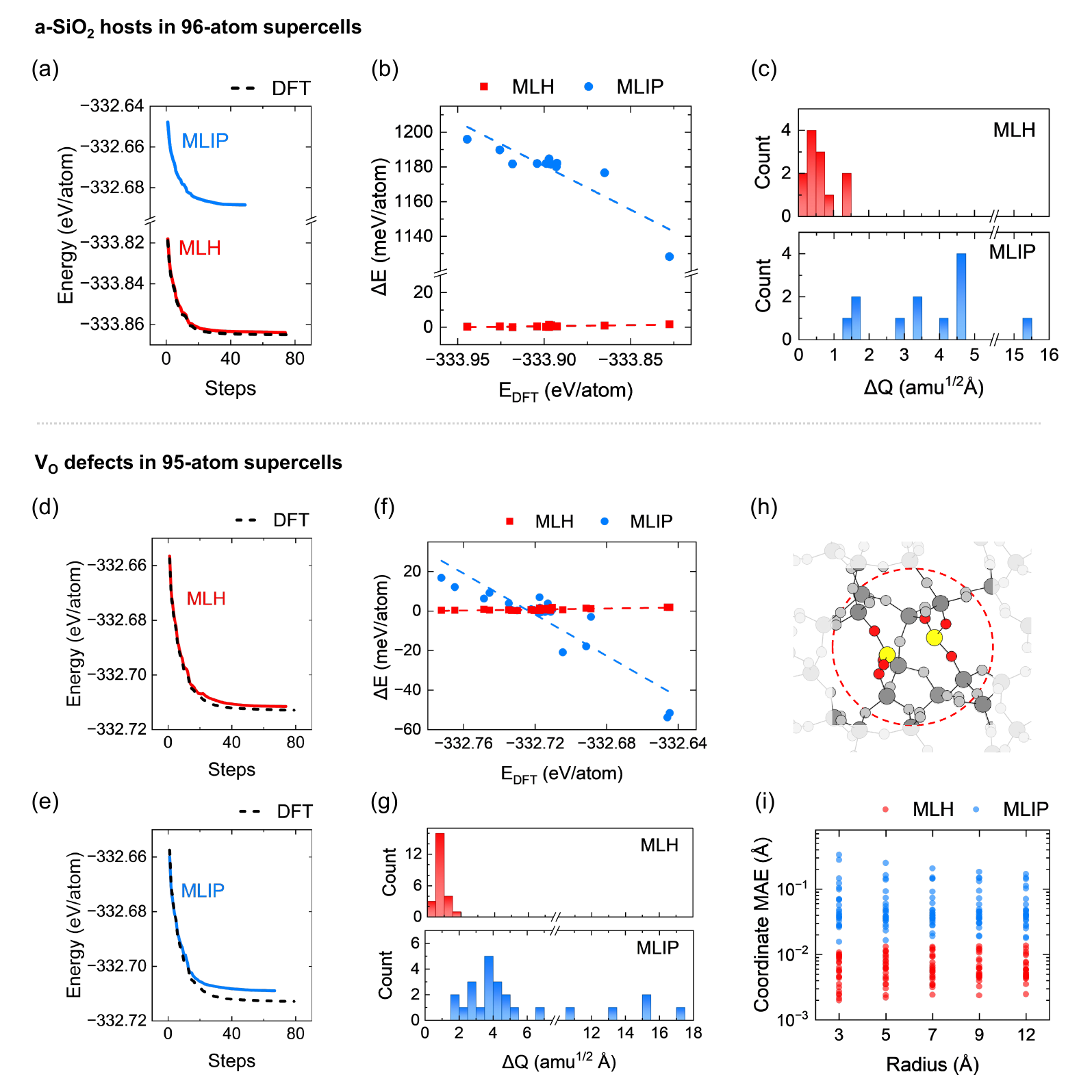}
    \caption{
    For a-SiO$_2$ hosts in 96-atom supercells: 
    (a) Representative total energy evolution during structural relaxation obtained with DFT, MLIP, and MLH.
    (b) Total energy errors of MLH and MLIP relative to DFT. Dashed lines represent linear fits to the respective data.
    (c) Histograms of $\Delta Q$ for MLH‑ and MLIP‑relaxed structures.
    For V$_\text{O}$ defects in 95-atom supercells: 
    Representative total energy evolution during structural relaxation, 
    (d) DFT vs. MLH, (e) DFT vs. MLIP.
    (f) Total energy errors of MLH and MLIP relative to DFT. Dashed lines represent linear fits to the respective data.
    (g) Histograms of $\Delta Q$ for MLH‑ and MLIP‑relaxed structures.
    (h) Illustration of the coordinate MAE for V$_\text{O}$ defects, computed for atoms within the cutoff radius (red circle) around the defect site.
    (i) Coordinate MAEs as a function of radius from the V$_\text{O}$ site, comparing MLH and MLIP.
    }
    \label{MLH_96}
\end{figure*}

To evaluate the scalability of the MLH model across different supercell sizes, we generate a series of a-SiO$_2$ host supercells with the number of atoms ranging from 96 to 576 and introduce a V$_\text{O}$ defect in each supercell. The lattice constants and radial distribution functions of these host amorphous structures are detailed in the Supplementary Material. Using the trained MLH model, we predict the Hamiltonian matrices for these systems and calculate their total energies and atomic forces. Meanwhile, DFT calculations are performed as references to evaluate errors and benchmark the computational efficiency of the MLH model. As shown in Fig.~\ref{MLH_size}(c), despite the training set containing only V$_\text{O}$ defect configurations within 95-atom supercells, the MLH model maintains stable predictive capability for defect and host systems in larger supercells, with energy errors below 1.2 meV/atom and atomic force MAEs of approximately 25 meV/Å for the majority of cases. The scalability of Hamiltonian and charge density predictions is discussed in the Supplementary Material. From the perspective of computational efficiency, the DFT calculation time increases sharply when the number of atoms exceeds 300 as presented in Fig.~\ref{MLH_size}(d), whereas that of the MLH model grows approximately linearly with the system size, demonstrating its significant efficiency advantage for the large-scale simulation of defects in complex materials.

\begin{figure*}[t!]
    \centering 
    \includegraphics[scale = 0.38]{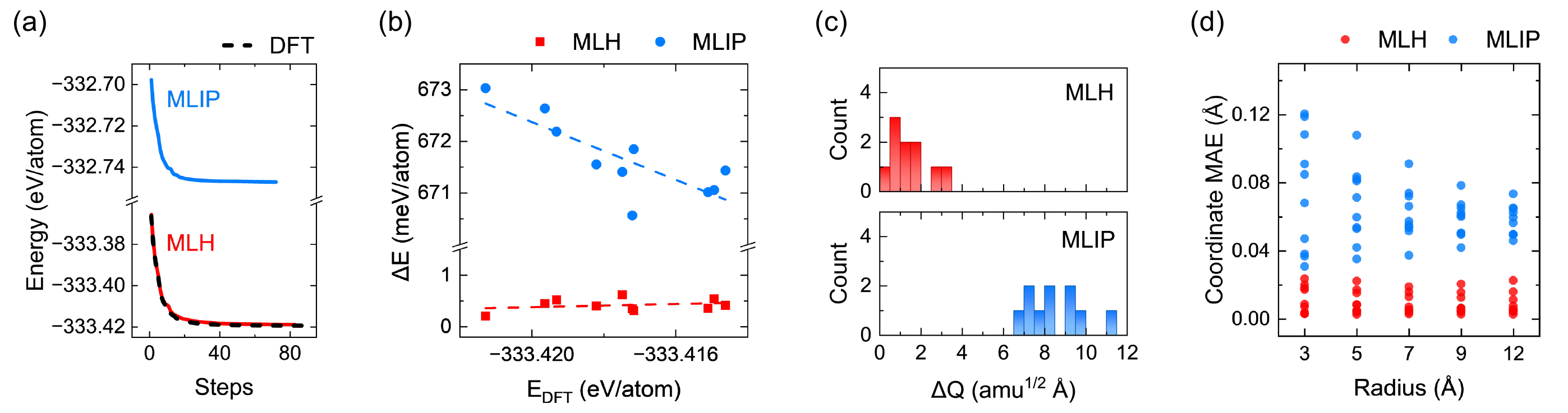}
    \caption{ 
    For V$_\text{O}$ defects in 215-atom supercells: 
    (a) Representative total energy evolution during structural relaxation using DFT, MLIP, and MLH.
    (b) Total energy errors of MLH and MLIP relative to DFT. Dashed lines represent linear fits to the respective data.
    (c) Histograms of $\Delta Q$ for MLH‑ and MLIP‑relaxed structures.
    (d) Coordinate MAEs, comparing MLH and MLIP.
    }
    \label{MLH_216}
\end{figure*}

\subsection{Structure relaxation: MLH vs. MLIP}

Leveraging the MLH model's accurate prediction capability for total energy and atomic forces, we subsequently apply it to accelerate structural relaxations of defect and host supercells. Using the 96-atom a-SiO$_2$ host supercell described in the previous section as a representative example, we perform the structural relaxation with the MLH, MLIP, and DFT methods. The relaxation trajectories are compared in Fig.~\ref{MLH_96}(a). The results demonstrate that the MLH model reproduces the relaxation trajectory of DFT, whereas the MLIP exhibits systematic energy errors, with predicted results significantly higher than those of DFT. To evaluate generalizability, we perform structural relaxations for twelve different 96-atom a-SiO$_2$ host supercells, which share the same lattice constants as the training set but possess distinct amorphous configurations. As shown in Fig.~\ref{MLH_96}(b), the MLIP exhibits a systematic overestimation of approximately 1180 meV/atom. This phenomenon is consistent with the systematic energy errors identified by Mosquera-Lois \textit{et al.}~\cite{mosquera2025point}. Additionally, we find that these MLIP errors vary with the total energy of the system, which is attributed to the softening of the potential energy surface~\cite{deng2025systematic}. 
This energy dependence is observed in the MLIP-predicted energies for both host supercells and defect supercells. A detailed discussion of this phenomenon is provided in subsequent sections.

In contrast to the MLIP predictions, the MLH model consistently predicts the total energies slightly above the DFT results, with a mean error of only 0.66 meV/atom. There are two main factors responsible for this slight energy elevation. First, MLH-relaxed structures are slightly different from those obtained by DFT. Second, in the KS framework, the ground-state total energy is determined by the ground-state Hamiltonian; however, the Hamiltonian predicted by MLH still deviates slightly from the exact one, which likewise yields a total energy higher than the DFT value.

To quantitatively evaluate the error of relaxed structures, we use the deviation of the generalized configuration coordinate ($\Delta Q$) to describe the configuration differences, which is defined as~\cite{alkauskas2014first}:
\begin{equation}
    \Delta Q = \sqrt{\sum_{\alpha, t} m_\alpha \Delta R_{\alpha,t}^2}
\end{equation}
where $\Delta R_{\alpha,t}$ denotes the coordinate deviation of atom $\alpha$ along the direction $t=\{x,y,z\}$, and $m_{\alpha}$ is the mass of the corresponding atom. Using the DFT-relaxed structures as references, the $\Delta Q$ values for MLH-relaxed structures are below 2 amu$^{1/2}$\,\AA, whereas for the majority of MLIP-relaxed structures, the $\Delta Q$ values range from 1 to 5 amu$^{1/2}$\,\AA, as shown in Fig.~\ref{MLH_96}(c). These findings reveal that even with only defect configurations in the training set, the MLH model can describe the interactions between atoms in the host environment, resulting in predictions that closely match DFT references.

Next, we introduce V$_\text{O}$ defects into these host supercells and examine the performance of the MLH model. All configurations encountered along the relaxation trajectories are excluded from the dataset. We compare the relaxation trajectories of DFT, MLIP, and MLH for one V$_\text{O}$ defect in a 95-atom supercell as an example. It is demonstrated in Fig.~\ref{MLH_96}(d) that the relaxation trajectory of MLH closely follows that of DFT, while Fig.~\ref{MLH_96}(e) reveals that the MLIP displays clearly larger energy deviations during structural relaxation than MLH does. Fig.~\ref{MLH_96}(f) presents the total energy errors of these relaxed V$_\text{O}$ defects. The results indicate that the MLIP suffers from potential energy surface softening. For high-energy configurations, such as defects with total energies exceeding -332.68 eV/atom, the MLIP underestimates the energies, resulting in negative errors; whereas for low-energy configurations, such as defect systems with energies around –332.76 eV/atom, the total energies are overestimated, yielding positive errors. In contrast to the MLIP, the MLH model remains in close agreement with DFT, yielding an average error of only 0.85 meV/atom. 

\begin{figure*}[t!]
    \centering 
    \includegraphics[scale = 0.62]{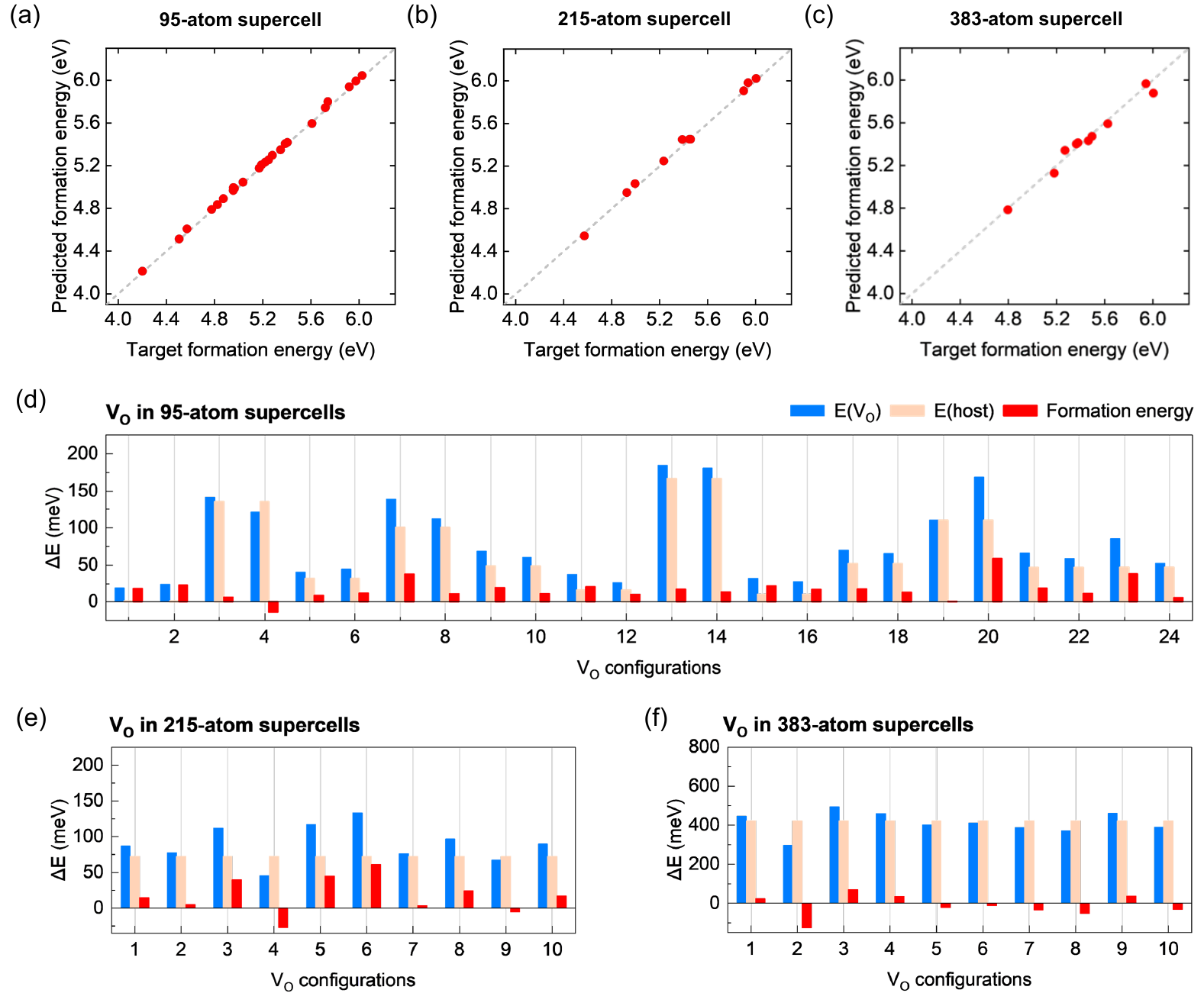}  
    \caption{(a)-(c) Comparisons of formation energies calculated by DFT and predicted by MLH for V$_\text{O}$ defects in 95-atom, 215-atom and 383-atom supercells.
    (d)-(e) Errors in defect total energy, corresponding host total energy, and formation energy relative to DFT values for V$_\text{O}$ defects within 95-atom, 215-atom and 383-atom supercells.
    }
    \label{MLH_Ef}
\end{figure*}

In terms of structural deviations, we calculate the $\Delta Q$ values for MLIP-relaxed and MLH-relaxed structures, as shown in Fig.~\ref{MLH_96}(g). It can be seen that all the $\Delta Q$ values of MLH-relaxed structures are below 2 amu$^{1/2}$\,\AA, whereas the $\Delta Q$ values for MLIP-relaxed structures range from 2 to 18 amu$^{1/2}$\,\AA{} with the peak value located at 4 amu$^{1/2}$\,\AA.
Considering that the structural relaxation for a defect supercell primarily affects atoms in the vicinity of the defect, we further calculate the coordinate MAE of atoms in a spherical region centered on the defect with radius $x$ ($x$= 3, 5, 7, 9, 12\,\AA)~\cite{yang2025modeling}, as depicted in Fig.~\ref{MLH_96}(h). We compare the coordinate MAEs between MLH-relaxed and MLIP-relaxed structures in Fig.~\ref{MLH_96}(i). The results show that the MAEs of the MLIP consistently exceed 0.01\,\AA, occasionally surpassing 0.1\,\AA, with large fluctuations. In contrast, the MAEs of the MLH model are mostly below 0.01\,\AA, an order of magnitude lower than that of the MLIP, and exhibit small variation over the entire radial range. This suggests a high agreement between the DFT-relaxed structures and MLH-relaxed structures.

For defect and host systems in larger supercells, the MLH model also demonstrates superior performance. 
Starting from a 216-atom a-SiO$_2$ host structure, we select ten oxygen atom sites to construct V$_\text{O}$ defects and perform structural relaxations for the host and defect configurations.
For the relaxed a-SiO$_2$ host, the MLH-predicted total energy deviates from DFT by only 0.34 meV/atom, and the MLH-relaxed structure yields a $\Delta Q$ of 1.3 amu$^{1/2}$\,\AA. In contrast, the MLIP gives a total energy error of 1195 meV/atom and a $\Delta Q$ of 8.4 amu$^{1/2}$\,\AA.
For V$_\text{O}$ defects, Fig.~\ref{MLH_216}(a) shows the relaxation trajectories of a V$_\text{O}$ defect as an example, and Fig.~\ref{MLH_216}(b) summarizes the total energy errors of these relaxed V$_\text{O}$ defects.
It can be seen that the MLIP also exhibits systematic energy errors and potential energy surface softening, with energy errors exceeding 670 meV/atom, whereas the MLH model achieves a mean error of only 0.41 meV/atom in total energy, highlighting its accurate and scalable predictive capability.
As illustrated in Fig.~\ref{MLH_216}(c) and Fig.~\ref{MLH_216}(d),  all the MLH-relaxed structures satisfy $\Delta Q < 4 $~amu$^{1/2}$\,\AA{} with most of the coordinate MAEs lower than 0.02\,\AA{} at different radii. However, the $\Delta Q$ values for MLIP-relaxed structures range from 6 to 12 amu$^{1/2}$\,\AA, and the coordinate MAEs reach 0.1\,\AA{} when the radius is under 5\,\AA.
This shows that, structural relaxations for hosts and defects in larger supercells can be accurately performed by the MLH model, even though it is trained only on defect configurations in small supercells.

\subsection{Formation Energy Calculations}

Given that the MLH-relaxed structures and the predicted total energies agree well with DFT, we proceed to calculate the formation energies of V$_\text{O}$ defects. Furthermore, we extend the analysis to a 383‑atom supercell to assess the model's performance. As shown in Fig.~\ref{MLH_Ef}(a-c), for V$_\text{O}$ defects in 95-, 215-, and 383-atom supercells, the predicted formation energies fall within the range of 4.2–6.0 eV, in good agreement with previously reported values~\cite{mukhopadhyay2005correlation,yue2017first,munde2017diffusion,gao2019mechanisms}.
The formation energy MAEs are 16, 26, and 45 meV, respectively, substantially lower than the errors reported in earlier studies~\cite{berger2025screening, song2023neural,chen2025simulating}.
It is worth noting that the formation energy errors are much lower than the MAEs of MLH-predicted total energies, which are approximately 80 meV (0.85 meV/atom), 100 meV (0.41 meV/atom) and 400 meV (1.1 meV/atom) for V$_\text{O}$ defects in 95-, 215-, and 383-atom supercells, respectively.
As illustrated in Fig.~\ref{MLH_Ef}(d) for the 14th V$_\text{O}$ configuration, the MLH model gives total energy errors of 180 meV for the defect and 166 meV for the host, with a formation energy error of only 14 meV. Similarly, the errors for V$_\text{O}$ defects in the 215-atom and 383-atom supercells are presented in Fig.~\ref{MLH_Ef}(e) and (f). These results indicate that, although MLH-predicted total energies exhibit errors for both host and defect systems, the formation energy errors are strongly reduced due to error cancellation. In contrast, MLIP shows systematic energy errors for host systems, which cannot be canceled by the errors in defect total energies. According to the results presented above, MLH's capability to deliver accurate formation energy predictions for defect systems in amorphous materials is confirmed.

Our results highlight the distinct advantages of the present MLH approach. Particularly when the computational cost of generating large supercell data becomes prohibitive, the MLH approach offers a highly practical alternative, enabling accurate defect property predictions without the need to explicitly train on large supercells. Beyond computational efficiency, it also provides detailed electronic structure information, including band structures and real-space wavefunctions of defect states. As illustrated for representative defects in the Supplementary Material, the ability to visualize both bonding and anti-bonding defect-state wavefunctions offers new physical insights into defect behavior in a-SiO$_2$.

\section{CONCLUSIONS}

In conclusion, we present an efficient and accurate approach for structural relaxation and formation energy prediction of defect supercells based on the MLH model. Taking V$_\text{O}$ defects in a-SiO$_2$ as a representative example, we constructed a dataset from DFT calculations, including SCF calculations for 120 V$_\text{O}$ configurations and structural relaxations for 12 V$_\text{O}$ configurations. Remarkably, even when trained solely on defect configurations in 95‑atom supercells, the MLH model accurately predicts total energies and atomic forces for both a-SiO$_2$ hosts and V$_\text{O}$ defects in larger supercells. The computational cost scales linearly with system size, improving simulation efficiency for large defect supercells. We further applied the MLH model to perform structural relaxations for a-SiO$_2$ hosts and V$_\text{O}$ defects, finding that the MLH-relaxed configurations and their total energies are in good agreement with DFT references, and the systematic energy errors observed in MLIPs are effectively avoided. In the calculation of the defect formation energies, errors in MLH-predicted total energies for host and defect systems largely cancel out, yielding accurate results with errors below 50 meV. These findings demonstrate that the proposed method offers an efficient and scalable pathway for large-scale defect simulations in complex materials.

\section{DATA AVAILABILITY}

The structures of a-SiO$_2$ hosts and V$_\text{O}$ defects, and the calculated energies supporting the key findings of this article are available within the article. All raw data of first-principles calculations are available from the corresponding authors upon reasonable request.

\begin{acknowledgments}

This work was supported by National Natural Science Foundation of China (12334005, 12174060, 12188101 and 12404089), 
National Key Research and Development Program of China (2022YFA1402904 and 2024YFB4205002), 
Science and Technology Commission of Shanghai Municipality (24JD1400600 and Explorer project 24TS1400500), and Project of MOE Innovation Platform.

\end{acknowledgments}

\bibliography{apssamp}

\end{document}